\documentclass{aa}
\usepackage{longtable}
\usepackage{natbib}
\usepackage{graphicx}
\usepackage[english]{babel}
\usepackage{lscape}

\newcommand{\gppr}{\stackrel{>}{\scriptstyle \sim}}
\newcommand{\gappr}{\raisebox{-0.4ex}{$\gppr$}}
\newcommand{\lppr}{\stackrel{<}{\scriptstyle \sim}}
\newcommand{\lappr}{\raisebox{-0.4ex}{$\lppr$}}

\newcommand{\Porb}{\mbox{$P_\mathrm{orb}$}}

\newcommand{\Mwd}{\mbox{$M_\mathrm{wd}$}}
\newcommand{\Msec}{\mbox{$M_\mathrm{sec}$}}

\newcommand{\Rsec}{\mbox{$R_\mathrm{sec}$}}

\newcommand{\Msun}{\mbox{$\mathrm{M}_{\odot}$}}
\newcommand{\Rsun}{\mbox{$R_{\odot}$}}
\newcommand{\Teff}{\mbox{$T_\mathrm{eff}$}}

\newcommand{\Lines}[3]{\Ion{#1}{#2}\,$\lambda\lambda$\,#3}
\newcommand{\Ion}[2]{#1{\,\scriptsize #2}}

\begin{document}

\title{Post common envelope binaries from SDSS. II:\\Identification of 9 close
binaries with VLT/FORS2\thanks{
Based on observations collected at the European Southern Observatory, 
Paranal, Chile under programme ID 078.D-0719.
This paper includes data gathered with the 6.5 meter Magellan Telescopes located at Las Campanas Observatory, Chile.}}
\authorrunning{Schreiber et al.}
\titlerunning{Identifying PCEBs with VLT/FORS2}
\author{M.R. Schreiber\inst{1}, 
B.T. G\"ansicke\inst{2}, 
J. Southworth\inst{2}, 
A.D. Schwope\inst{3},
D. Koester\inst{4}
}
\institute{
Departamento de F\'isica y Astronom\'ia, Facultad de Ciencias, Universidad 
de Valpara\'iso, Valpara\'iso, Chile 
\and
Department of Physics, University of Warwick, Coventry CV4 9BU, UK
\and
Astrophysikalisches Institut Potsdam, An der Sternwarte 16, D-14482
Potsdam, Germany
\and
Institut f\"ur Theoretische Physik und Astrophysik, University of Kiel,
24098 Kiel, Germany
\\
\email{Matthias.Schreiber@uv.cl}
}
\offprints{M.R. Schreiber}

\date{Received / Accepted }

\abstract {
Post common envelope binaries (PCEBs) consisting of a white
dwarf and a main sequence star are ideal systems to calibrate current
theories of angular momentum loss in close compact binary stars. The
potential held by PCEBs for further development of close binary
evolution could so far not be exploited due to the small number 
of known systems and inhomogeneity of the sample.  
The Sloan Digital Sky Survey is changing this scene
dramatically, as it is very efficient in identifying white dwarf/main
sequence (WDMS) binaries, including both wide systems whose stellar
components evolve like single stars and - more interesting in the
context of close binary evolution - PCEBs.}
{We pursue a large-scale follow-up survey to identify and characterise
the PCEBs among the WDMS binaries that have been found with SDSS. We
use a two-step strategy with the identification of PCEBs among WDMS binaries in
the first phase and orbital period determinations in the 
second phase. Here
we present first results of our ESO-VLT/FORS2 pilot-study that has the
target of identifying the PCEBs among the fainter 
($g\gappr18.5$) SDSS WDMS binaries.}
{From published SDSS catalogues we selected 26 WDMS binaries to be
observed with ESO-VLT/FORS2 in service mode.  The design of the
ESO-VLT/FORS2 observations was to get two spectra per object separated
by at least one night.  We used the \Lines{Na}{I}{8183.27,8194.81}
doublet to measure radial velocity variations of our targets and
performed additional follow-up spectroscopy using Magellan-Clay/LDSS3
of two systems showing significant radial velocity variations.  Using
a spectral decomposition/fitting technique we determined the white
dwarf effective temperatures and surface gravities, masses, and
secondary star spectral types for all WDMS binaries in our sample.  }
{Among the 26 observed WDMS binaries  we find $9$ strong PCEB candidates
showing clear ($\geq\,3\sigma$) radial velocity variations, and we
estimate the fraction of PCEBs among SDSS WDMS binaries to be
$\sim35\pm12\%$. We find indications for a dependence of the relative number 
of PCEBs among SDSS WDMS binaries on the spectral type of the secondary star. 
These results are subject to small number statistics and need to be confirmed
by additional observations. The orbital periods of two PCEB
candidates, SDSS\,J1047+0523 and SDSS\,J1414--0132, we measured
to be $9.17$\,hrs and $17.48$\,hrs respectively. }
{This pilot study demonstrates that our survey is highly efficient in
identifying PCEBs among the SDSS WDMS binaries, and will indeed
provide the observational parameters that are necessary to constrain
the theoretical models of close binary evolution.}


\maketitle

\section{Introduction}

The majority of all stars are born in binary or multiple star systems
and a significant fraction of these will interact at some point during
their lives. Once the more massive stars leaves the main sequence and
depending on the initial conditions, dynamically unstable mass
transfer or a tidal instability may force the system to enter the
common envelope (CE) 
phase \citep[see][for more details]{taam+ricker06-1}. 
In this phase a gaseous envelope around
the entire binary forms and friction within the envelope is
significantly reducing the binary separation. 
Since \citet{paczynski76-1} it is generally believed that 
virtually all close compact binary systems ranging
from X-ray binaries to double white dwarf binaries or cataclysmic
variables, to name a few, have formed through CE evolution. 
\citet{willems+kolb04-1} performed comprehensive binary
population synthesis 
studies for white dwarf/main sequence binaries (WDMS),
and find that 
dynamically stable mass transfer may produce some 
additional short orbital period ($\Porb\gappr\,10$\,days)
systems primarily consisting of low-mass He white dwarfs 
($0.1\leq\Mwd\leq0.4\Msun$) and relatively
massive secondary stars (up to $6\Msun$).
In the vast majority ($\sim75-99\,\%$) of cases, however, a CE phase is needed to produce short orbital period ($\Porb\lappr\,400$\,days) WDMS.

From theoretical simulations we learned that the CE phase is probably
very short, i.e. $\lappr10^3$\,yrs, that the spiral in starts rapidly
after the onset of the CE phase, and that the expected form of post-CE
planetary nebular are bipolar 
\citep[see e.g.][and references theirin]{morris81-1,taam+bodenheimer89-1,taam+sandquist00-1,webbink07-1}. Hydrodynamical studies have been
carried for a number of parameters \citep[e.g.][for a recent
review]{taam+ricker06-1},however, these expensive simulations cannot
be used in population synthesis models.  Instead simple equations
relating the total energy of the binary before and after the CE phase
are applied \citep[e.g][]{willems+kolb04-1}.  
Usually, the CE
phase is simply approximated by a parameterised energy
\citep{webbink84-1,willems+kolb04-1} or angular momentum equation
\citep{nelemansetal00-1}. Both descriptions differ significantly in
the predicted outcome of the CE phase, i.e. the energy equation
predicts a much stronger shrinkage than the angular momentum equation 
\citep{nelemans+tout05-1}. 
In addition, in both prescriptions the efficiency to ``use'' the
orbital energy (or angular momentum) to expel the envelope is
uncertain. Hence, the CE phase is probably the least understood part
of close binary evolution and currently intensively debated (for
recent discussions of the CE phase see
\citet{nelemans+tout05-1,webbink07-1,beeretal07-1}).

\begin{figure*}
\includegraphics[angle=-90,width=0.49\textwidth]{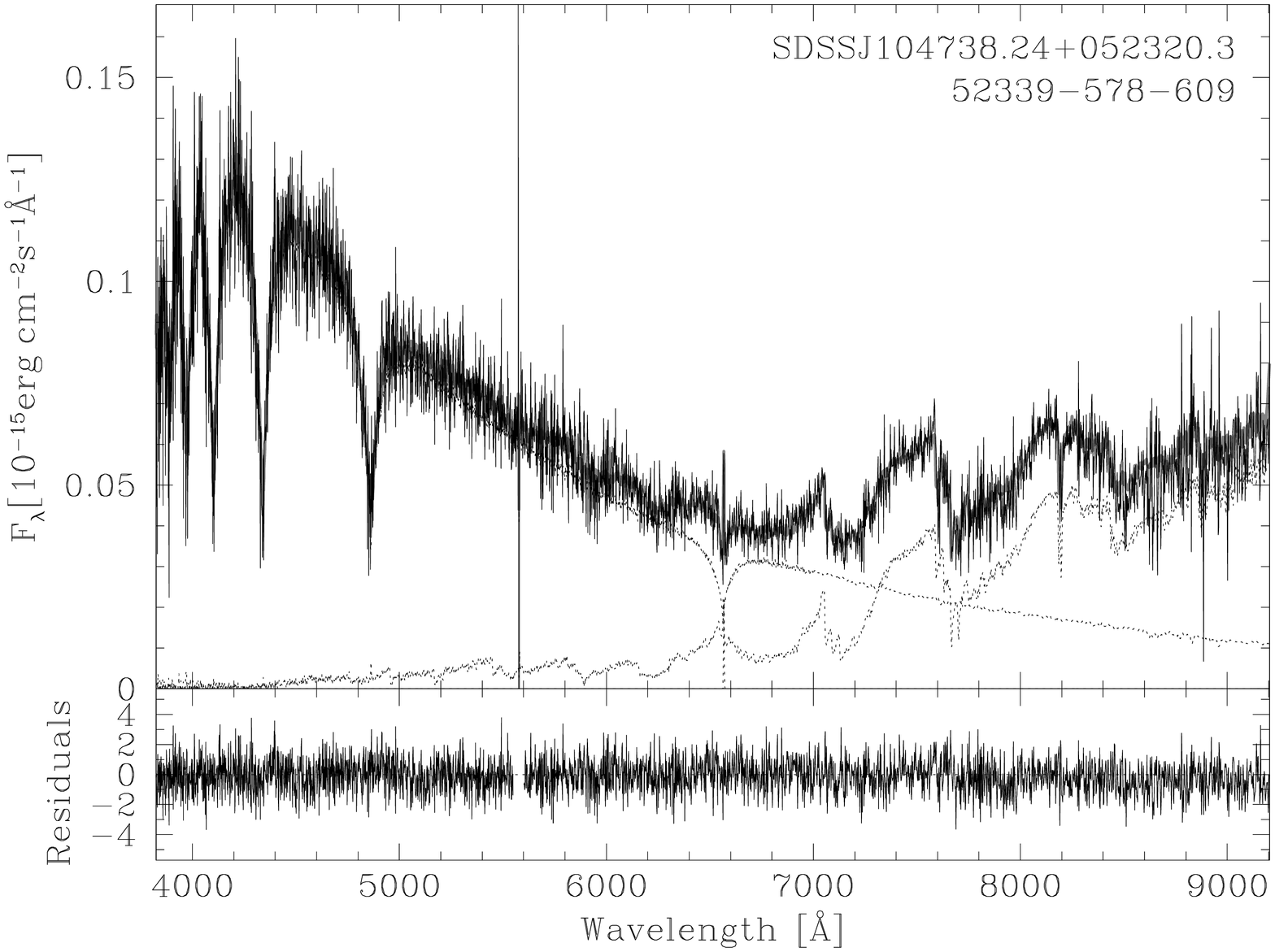}
\hfill
\includegraphics[angle=-90,width=0.49\textwidth]{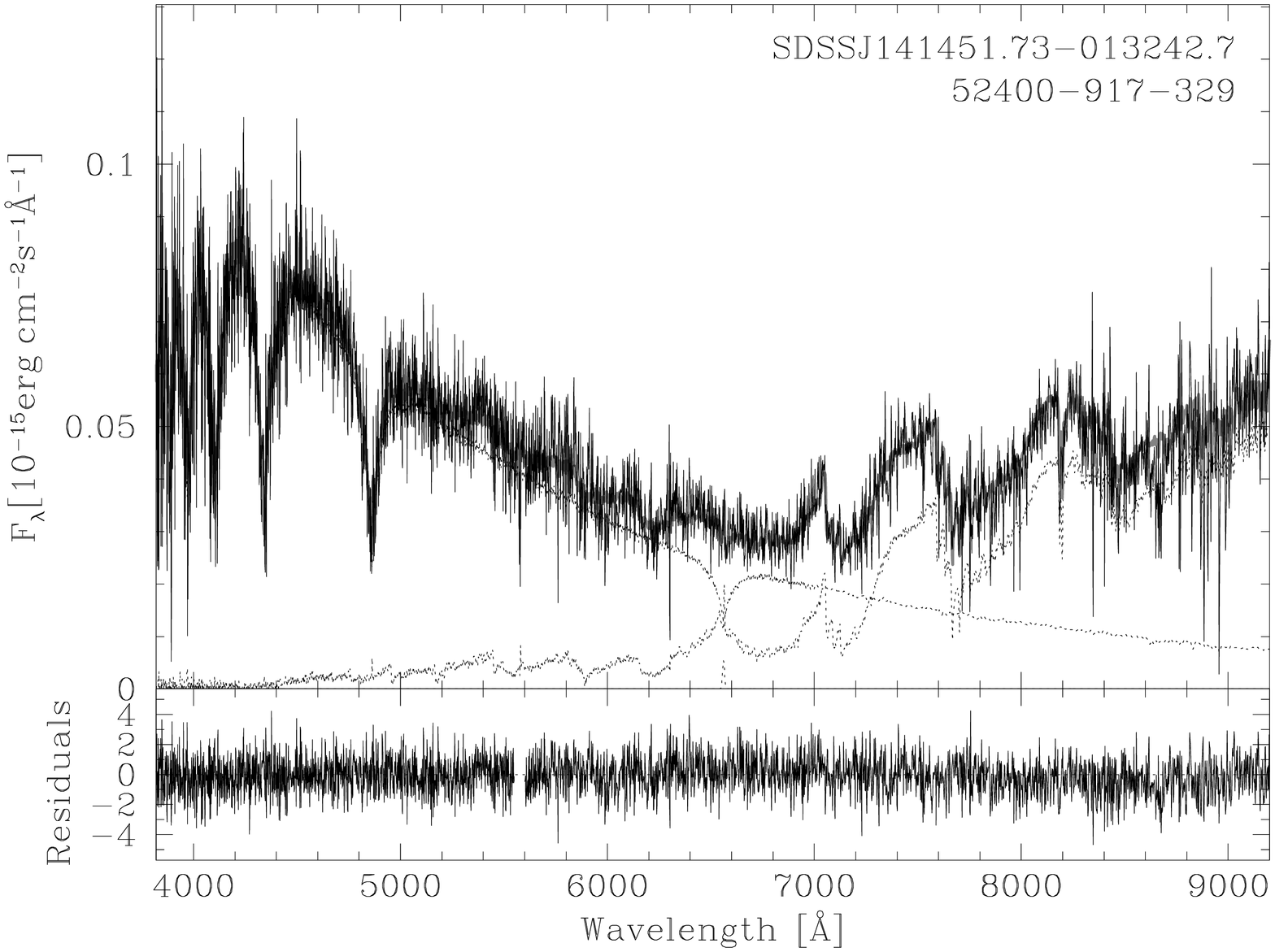}
\caption{
\label{f-compositefit}
Two-component fits to the 2 PCEBs SDSS\,J1047+0523 and
SDSS\,J1414--0132, for which we determined orbital periods from
Clay/LDSS3 spectroscopy. Top panels: the WDMS spectra (black lines)
along with the white dwarf and M-dwarf templates (dotted lines). 
The MJD--PLT--FIB identifiers of the two SDSS spectra are also given.
Bottom panels: residuals from the fit.}
\end{figure*}

Once the envelope is expelled, the evolution of the PCEB is mainly
driven by angular momentum loss (AML) due to magnetic braking.
Unfortunately, the two currently favoured prescriptions for magnetic
braking \citep{verbunt+zwaan81-1, andronovetal03-1}, differ by up to
two orders of magnitude in the predicted angular momentum loss rate.  
Even worse, it is not clear whether magnetic
braking is present in all PCEBs, or whether it terminates in systems
where the secondary star is fully convective. In order to explain the
orbital period gap observed in the period distribution of CVs, one in
fact needs to assume the latter, known as ``disrupted magnetic
braking'', \citep[e.g.][]{king88-1,howelletal01-1}. In contrast to
this, observations of single low mass stars do not show any evidence
for such a discontinuity \citep[e.g.][]{pinsonneaultetal02-1}.

Significant progress in our understanding of the CE phase and AML in
PCEBs certainly requires both continuous theoretical efforts and
innovative observational input.  The potential that a large sample of
white dwarf plus main sequence PCEBs could provide in this context has
been first realised by \citet{schreiber+gaensicke03-1}.  They showed
that it is possible to determine the orbital period the PCEB had 
when it left the CE phase and the white dwarf temperature as well as 
the orbital period it will have when entering the next
phase of interaction (becoming a cataclysmic variable). More recently,
\citet{politano+weiler06-1} showed that the disrupted magnetic braking
hypothesis can be tested by knowing the relative number of PCEBs among
white dwarf/main sequence (WDMS) binaries as a function of secondary
spectral type. Finally \citet{schreiberetal07-1} recently summarised
that a representative sample of PCEBs could probably constrain the CE
efficiency, the description of the CE phase, and the strength of
magnetic braking.

Unfortunately, only a very limited number of PCEBs with known
effective temperature and orbital period is currently available.  The
PCEB sample that \citet{schreiber+gaensicke03-1} analysed consisted of
only 30 well-studied short-period WDMS known in
2002. Since then, e.g. \citet{gaensickeetal04-1},
\citet{morales-ruedaetal05-1}, \citet{vandenbesselaaretal07-1},
\citet{shimanskyetal06-1}, \citet{aungwerojwitetal07-1}, and
\citet{tappertetal07-1} added additional systems and the current
number of well studied PCEBs is close to $50$.  The number of systems
required to constrain close binary evolution is certainly much
higher. \citet{schreiberetal07-1} estimate that about $200$ PCEBs with
known white dwarf temperatures, secondary star parameters, and orbital
periods are needed.

\begin{figure*}
\includegraphics[width=0.49\textwidth]{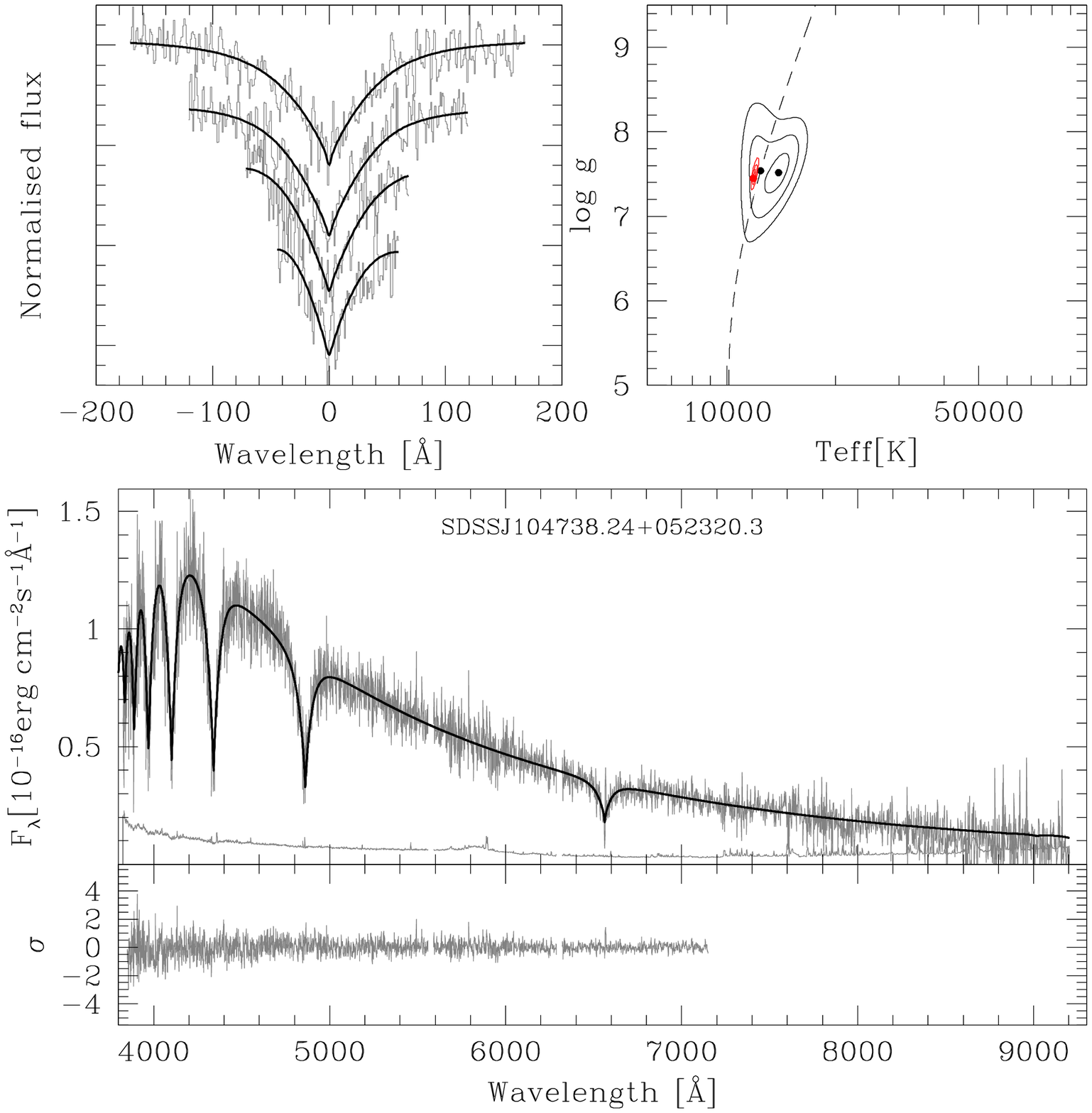}
\hfill
\includegraphics[width=0.49\textwidth]{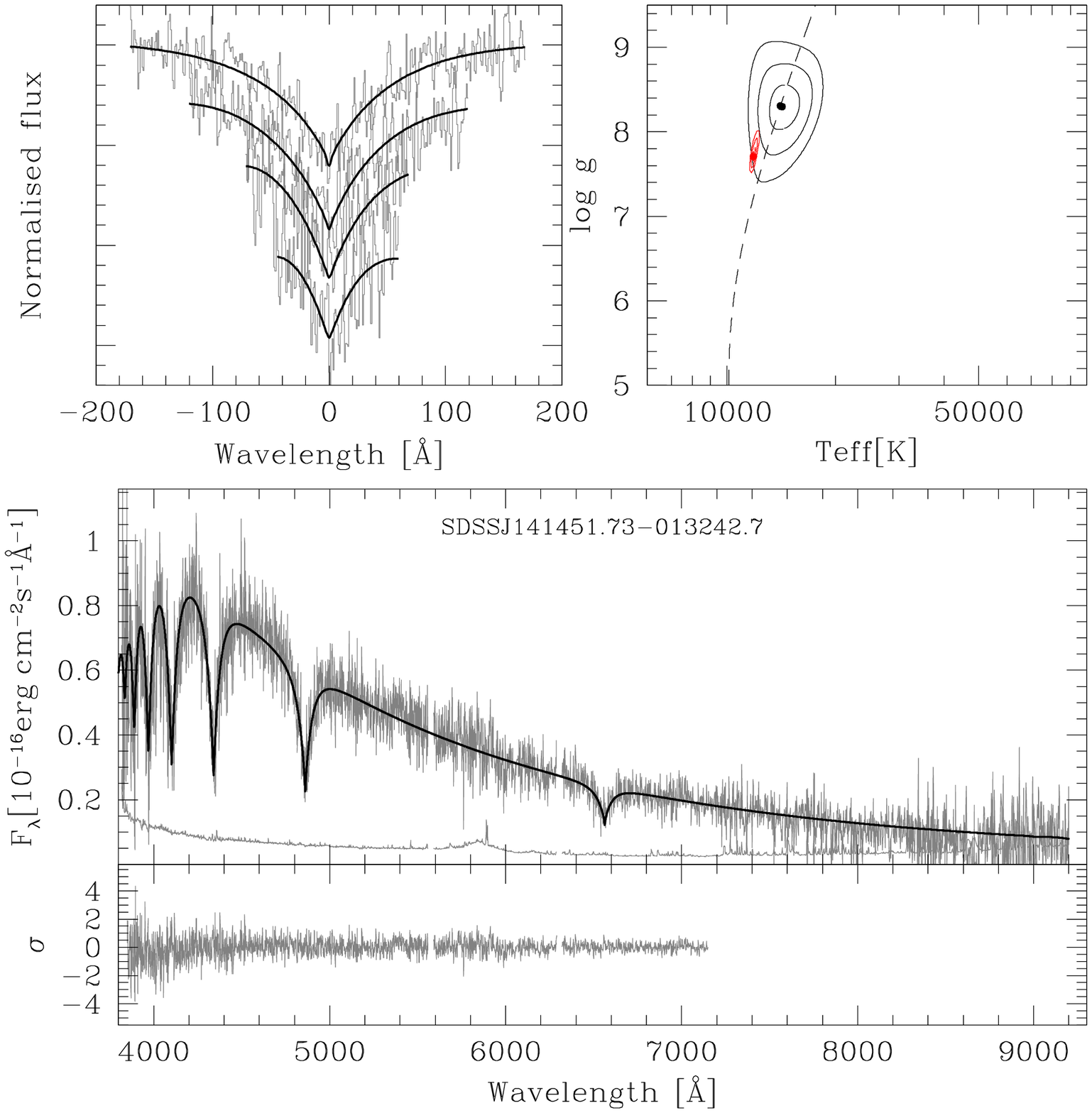}
\caption{
\label{f-wdfit}
Spectral model fits to the white dwarf components of 
  SDSS\,J1047+0523 and SDSS\,J1414--0132, obtained by subtracting the
  best-fit M-dwarf template from their SDSS spectra. Top left panels:
  normalised H$\beta$ to H$\epsilon$ line profiles (top to bottom,
  gray lines) along with the best-fit white dwarf model (black
  lines). Top right panels: 1, 2, and 3$\sigma$ $\chi^2$ contour
  plots in the $\Teff-\log g$ plane. The black contours refer to the
  best line profile fit, the red contours to the fit of the whole
  spectrum. The dashed line indicates the location of maximum H$\beta$
  equivalent width in the $\Teff-\log g$ plane, dividing it into
  ``hot'' and ``cold'' solutions.  The best-fit parameters of the
  ``hot'' and ``cold'' normalised line profile solutions and of the
  fit to the $3850-7150$\,\AA\ range are indicated by black and red
  dots, respectively. Bottom panels: the white dwarf spectrum and
  associated flux errors (gray lines) along with the best-fit white
  dwarf model (black lines) to the $3850--7150$\,\AA\ wavelength range
  (top) and the residuals of the fit (gray line, bottom).}
\end{figure*}

The Sloan Digital Sky Survey (SDSS) is currently opening a new
era of observational population studies for a wide variety of
astronomical objects, including WDMS binaries. As by-product of the
quasar search within SDSS, more than 1000 WDMS binaries have been
identified already within Data Release 5 \citep{eisensteinetal06-1,
silvestrietal07-1, southworthetal07-2}. Though very large, this set of
WDMS binaries is still subject to selection effects, as it favours
systems with bluish colours, i.e. hot white dwarfs. To compensate for
this, we have initiated a dedicated WDMS identification programme as
part of SDSS\,II-SEGUE, which targets systems with a strong
contribution of the companion star \citep[see][]{schreiberetal07-1}.
In parallel, we have started the identification of PCEBs among the
WDMS binaries identified by SDSS by means of a radial velocity
survey.  \citeauthor{rebassa-mansergasetal07-1}
(\citeyear{rebassa-mansergasetal07-1}, henceforth Paper\,I) presented
a first set of 18 PCEB candidates from radial velocity variations
detected among multiple SDSS spectra.  In this paper we present
VLT/FORS2 observations of 26 faint ($g\gappr18.5$) WDMS
binaries which led to the identification of 9 new PCEBs through radial
velocity variations. The orbital periods of two of these PCEBs were
determined from additional Magellan-Clay/LDSS3 observations. The
structure of the paper is as follows.  In Section\,\ref{s-modelling}
we describe the selection of the 26 VLT targets and the
determination of their stellar parameters. Details on the observations
and data reduction of the VLT and Magellan-Clay runs are given in
Sect.\,\ref{s-obs}. After describing our method of measuring
radial velocities and determining orbital periods (Sect.\,\ref{s-rv})
we discuss our results in the context of our large-scale PCEB survey.

\section{\label{s-modelling}Target Selection and Stellar parameters}
For this project, we selected 26 WDMS binaries with $i<19.5$ and
visible from Paranal from the SDSS Data release 5 (DR5,
\citealt{adelman-mccarthyetal07-1}), using the lists compiled by
\citet{eisensteinetal06-1, silvestrietal07-1, southworthetal07-2}.
Given the recently suggested test of disrupted magnetic braking
\citep{politano+weiler06-1}, we focussed this selection on WDMS
binaries with intermediate M-type (M2--M5)
companions. Table\,\ref{t-targets} lists the SDSS object names and the
$i$ magnitudes of our targets.

We have determined the stellar parameters of these 26 WDMS
binaries following the method described in Paper\,I, with the
only difference of now using spectra retrieved from Data Release 6,
which have undergone a complete re-reduction that substantially
improves the flux calibration \citep{adelman-mccarthyetal08-1}.  For
the sake of brevity, we provide here only a brief summary of the
adopted procedures, for full details the reader is referred to
Paper\,I.  We first carry out a two-component fit to the SDSS WDMS
spectra using libraries of observed white dwarf and M-dwarf template
spectra, normalised to stellar surface fluxes. These template
libraries are also compiled from SDSS\,DR6. This stage
provides us with a spectral type for the companion star and a flux
scaling factor for the M-dwarf template which, together with a
spectral type-radius relation, yields a distance estimate. In a second
step, we subtract the best-fit M-dwarf template from the WDMS spectra,
and fit the residual white dwarf spectra with a standard $\chi^2$
method for variable white dwarf effective temperature (\Teff) and
surface gravity ($\log g$), using a grid of pure-hydrogen (DA) model
spectra computed with the code of \citet{koesteretal05-1}\footnote{We
limit the spectral fitting to DA white dwarfs, but provide a spectral
type of the companion for the two helium atmosphere (DB) WDMS binary
in our sample.}.  Two flavours of white dwarf fits were carried out
for each object: (1) a fit to the normalised line profiles. Due to the
degeneracy in Balmer line equivalent width as a function of
temperature two acceptable fits are usually found, a ``hot'' and a
``cold'' solution. (2) a fit to the continuum plus lines over the
range $3850-7150$\,\AA\,. The parameters from this fit are subject to
substantial systematic uncertainties due to inaccuracies in the
\textit{spectral slope}, and we only use them to break the degeneracy
between the ``hot'' and ``cold'' line profile solutions. Examples of
the spectral decomposition and of the white dwarf fits are shown in
Fig.\,\ref{f-compositefit} and Fig.\,\ref{f-wdfit}.

\begin{figure}
\centerline{\includegraphics[width=\columnwidth]{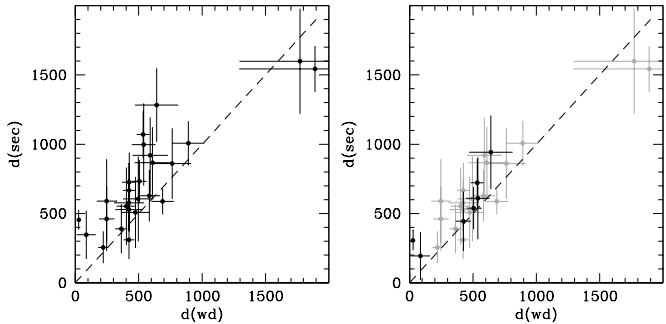}}
\caption[]{\label{f-distdist} Distances of the 24 WDMS binaries in
  Table\,\ref{t-targets} that contain a DA white dwarf. The spectral
  modelling (Sect.\,\ref{s-modelling}) provides two independent
  distance estimates for the individual stellar components. Nine
  systems have $d_\mathrm{sec}>d_\mathrm{wd}$ by more than
  $1\,\sigma$. Adopting secondary star spectral types later by one
  class brings most of them onto the $d_\mathrm{sec}=d_\mathrm{wd}$
  relation. An earlier appearance of the spectral type determined from
  an optical spectrum may be related to stellar activity. See Paper\,I
  for additional discussion.}
\end{figure}

We then used an updated version of the tables in
\citet{bergeronetal95-2}\footnote{http://www.astro.umontreal.ca/~bergeron/CoolingModels/},
which incorporate the structure and evolution calculations of
\citet{wood95-1} and \citet{fontaineetal01-1}, to compute white dwarf
masses and radii from \Teff\ and $\log g$. Finally, the white dwarf
radii and flux scaling factors of the white dwarf models yield a
second distance estimate. Statistical errors on the different
properties are propagated through this process, but should be
considered with some caution as additional systematic uncertainties
are not easy to quantify. The resulting stellar parameters are listed
in Table\,\ref{t-targets}. Inspection of
Fig.\,\ref{f-distdist} (left panel) shows that in about one third of
the systems the distance estimate based on the fit of the secondary
star exceeds that based on the white dwarf fit by a significant
($>1\,\sigma$) factor. A similar ratio of systems with
$d_\mathrm{sec}>d_\mathrm{wd}$ has been found in the much larger
sample analysed in Paper\,I, which also discusses in detail possible
explanations for this finding. Our favoured hypothesis from Paper\,I
is that stellar activity causes large scatter in the spectral
type-radius relation of low-mass stars (in fact, larger than the
scatter in their mass-radius relation that is subject to thorough
investigations, e.g. \citealt{ribas06-1,lopez-morales07-1}), with a
tendency for spectral types too early for their radius. In the
right-hand panel of Fig.\,\ref{f-distdist}, we show that adopting
secondary stars of one spectral class later than in
Table\,~\ref{t-targets} brings the nine systems which had
$d_\mathrm{sec}>d_\mathrm{wd}$ on, or close to, the
$d_\mathrm{sec}=d_\mathrm{wd}$ relation.

A final note concerns SDSS\,J2324--0931, which contains a DB
white dwarf. The DR3 SDSS spectrum of SDSS\,J2324--0931 was analysed
by \citet{vandenbesselaaretal05-1} who adopted a method similar to our
spectral decomposition, i.e. fitting the WDMS spectrum with the
combination of observed white dwarf and M-dwarf templates. Different
to our approach, they fixed the system distance to be the same for
both stars. Their analysis resulted in $\Teff=36815$, Sp(2)\,=\,M3V,
and $d=911$\,pc. Our spectral decomposition results in Sp(2)\,=\,M4V
and $d=633\pm188$\,pc. Forcing the spectral type to M3V, we obtain a
distance of $879\pm140$\,pc. Given that the typical uncertainty in the spectral
type is one spectral class, the two results are hence consistent.

\begin{table*}
\caption{\label{t-targets}Stellar parameter of the 26 WDMS binaries
  observed with FORS2, as determined from spectral modelling
  (Sect.\,\ref{s-modelling}). Upper limits on the orbital periods for
  the nine PCEBs identified by our VLT/FORS2 observations are
  calculated as described in Sect.\,\ref{s-porblimits}}
  \setlength{\tabcolsep}{1ex}
\begin{tabular}{cccccccccc}
\hline\noalign{\smallskip}
System (SDSS\,J) & MJD-plate-fibre & i & $T$[K] & $\log g$ & $M_\mathrm{wd}$[\Msun] & $d_\mathrm{wd}$[pc]  & Sp  & $d_\mathrm{sec}$[pc] & notes\\
\noalign{\smallskip}\hline\noalign{\smallskip}
%
%
001726.63--002451.1 & 52559-1118-280 & 18.3 & 12828$\pm$ 2543 & 8.00$\pm$0.46 & 0.61$\pm$0.29 &  423$\pm$119 & 4 &  577$\pm$171   \\
--                  & 52518-0687-153 & 18.3 & 13588$\pm$ 1803 & 8.11$\pm$0.39 & 0.68$\pm$0.25 &  423$\pm$107 & 4 &  529$\pm$157   \\
022835.92--074032.3 & 51908-0454-534 & 18.7 & 13588$\pm$  948 & 7.76$\pm$0.17 & 0.48$\pm$0.09 &  423$\pm$ 43 & 6 &  311$\pm$140 & \\
032038.72--063822.9 & 51924-0460-432 & 18.7 & 11045$\pm$  479 & 8.49$\pm$0.33 & 0.91$\pm$0.19 &  247$\pm$ 61 & 5 &  461$\pm$232 & 1, $\Porb\le143.1$\,d \\
081959.20+060424.2  & 52962-1296-138 & 19.2 & 15782$\pm$ 1422 & 7.67$\pm$0.34 & 0.45$\pm$0.17 &  764$\pm$150 & 4 &  860$\pm$255   \\
083354.84+070240.1  & 52963-1297-511 & 18.5 & 16336$\pm$  730 & 8.03$\pm$0.16 & 0.63$\pm$0.10 &  424$\pm$ 45 & 4 &  666$\pm$198 & 1, $\Porb\le14.7$\,d\\
085542.49+044717.7  & 52670-1190-512 & 19.1 & 10073$\pm$  336 & 8.46$\pm$0.43 & 0.89$\pm$0.25 &  247$\pm$ 77 & 5 &  589$\pm$297 & 1, $\Porb\le61.9$\,d\\
092451.63--001736.4 & 52000-0474-175 & 17.6 & 24163$\pm$ 1359 & 7.82$\pm$0.20 & 0.54$\pm$0.11 &  688$\pm$ 89 & 3 &  588$\pm$ 94   \\
093904.03+051114.8  & 52710-0993-296 & 19.2 & 12681$\pm$ 1406 & 7.69$\pm$0.42 & 0.44$\pm$0.23 &  591$\pm$139 & 4 &  919$\pm$273   \\
095108.74+025507.5  & 51908-0481-419 & 18.5 & 13432$\pm$ 1007 & 7.80$\pm$0.22 & 0.50$\pm$0.12 &  425$\pm$ 57 & 4 &  726$\pm$215   \\
100953.69--002853.4 & 51909-0270-181 & 19.0 & 21535$\pm$ 1927 & 8.35$\pm$0.32 & 0.84$\pm$0.19 &  609$\pm$141 & 4 &  867$\pm$257   \\
104358.59+060320.9  & 52643-1000-055 & 18.6 & 11565$\pm$  536 & 8.39$\pm$0.24 & 0.85$\pm$0.15 &  220$\pm$ 39 & 6 &  256$\pm$115   \\
104738.24+052320.3  & 52339-0578-609 & 18.7 & 12392$\pm$ 1847 & 7.54$\pm$0.42 & 0.38$\pm$0.20 &  474$\pm$111 & 5 &  508$\pm$256 & 1, 2, $\Porb<0.8$\,d\\
110151.34+122241.5  & 53119-1603-492 & 18.8 & 20331$\pm$  817 & 7.83$\pm$0.15 & 0.53$\pm$0.08 &  534$\pm$ 50 & 3 & 1071$\pm$171   \\
112623.87+010856.8  & 51614-0281-402 & 17.5 & 16717$\pm$  781 & 7.91$\pm$0.18 & 0.57$\pm$0.10 &  505$\pm$ 57 & 2 &  733$\pm$152   \\
122208.48+093406.2  & 52672-1230-73  & 18.3 & 30419$\pm$ 2053 & 7.34$\pm$0.49 & 0.38$\pm$0.18 & 1891$\pm$599 & 0 & 1543$\pm$166 & 4  \\ 
125250.03--020608.1 & 51694-0338-343 & 18.5 & 22037$\pm$ 2333 & 8.29$\pm$0.38 & 0.80$\pm$0.23 &  641$\pm$171 & 2 & 1282$\pm$265   \\
141451.73--013242.7 & 52400-0917-329 & 19.0 & 14065$\pm$ 1452 & 8.31$\pm$0.27 & 0.80$\pm$0.17 &  402$\pm$ 77 & 5 &  552$\pm$278 & 1, 3, 6, $\Porb<3.4$\,d\\
142541.32+023047.4  & 51999-0535-348 & 18.9 & 13127$\pm$ 1014 & 7.91$\pm$0.22 & 0.56$\pm$0.13 &  362$\pm$ 48 & 6 &  390$\pm$176   \\
150657.58--012021.7 & 52426-0922-536 & 18.4 & 15782$\pm$ 1005 & 7.72$\pm$0.24 & 0.47$\pm$0.13 &  584$\pm$ 85 & 4 &  629$\pm$187 & 1, $\Porb<9.1$\,d \\
204541.90--050925.7 & 52145-0635-439 & 18.7 & 26801$\pm$ 1432 & 7.78$\pm$0.22 & 0.53$\pm$0.11 &  892$\pm$127 & 3 & 1007$\pm$160 & 1, $\Porb<49.1$\,d \\
205059.37--000254.3 & 52466-0982-567 & 19.2 & 18330$\pm$ 1169 & 8.27$\pm$0.24 & 0.78$\pm$0.15 &  539$\pm$ 93 & 4 &  998$\pm$296   \\
212320.74+002455.5  & 52523-0987-484 & 19.2 & 12250$\pm$ 1794 & 7.47$\pm$0.42 & 0.35$\pm$0.19 &  496$\pm$119 & 5 &  606$\pm$305 & 1, $\Porb<0.4$\,d \\
215614.57--000237.4 & 52078-0371-025 & 18.5 &  7910$\pm$  926 & 9.50$\pm$0.74 & 1.37$\pm$0.21 &   86$\pm$ 76 & 5 &  346$\pm$174 & 1, 4, 6, $\Porb<44.0$\,d\\
220338.61--001750.7 & 52173-0372-034 & 18.7 & 32972$\pm$ 1703 & 7.24$\pm$0.38 & 0.37$\pm$0.11 & 1772$\pm$477 & 1 & 1598$\pm$379   \\
231221.59+010127.0  & 51811-0381-570 & 17.3 &              -- &            -- &            -- &           -- & 3 &  455$\pm$ 72 & 5\\
232438.31--093106.4 & 52203-0645-637 & 18.4 &              -- &            -- &            -- &           -- & 4 &  633$\pm$188 & 5\\
\noalign{\smallskip}\hline\noalign{\smallskip}
\multicolumn{10}{p{\textwidth}}{(1) PCEB candidate on the base of
radial velocity variations detected among the available FORS2 and SDSS
spectra; (2) $\Porb=9.17$\,h determined from Magellan-Clay/LDSS3
spectroscopy; (3) $\Porb=17.48$\,h determined from
Magellan-Clay/LDSS3 spectroscopy; (4) blue component is faint, white
dwarf parameter uncertain; (5) DB white dwarf; (6) poor quality spectrum.} \\
\end{tabular}
\end{table*}

\section{\label{s-obs}Observations}
\subsection{VLT/FORS2}
Intermediate resolution spectroscopy of 26 WDMS binaries were
obtained between August 16, 2007 and March 26, 2007 with FORS2 on the
VLT/UT1 (Antu). The exposure time was 900\,sec for all objects. The
observations were carried out using the 1028z grism and a 1\arcsec\
slit, resulting in a spectral coverage of 7830--9570\,\AA. The images
were processed using the {\sc starlink} packages {\sc figaro} and {\sc
kappa}, and the spectra were optimally extracted \citep{horne86-1}
using the {\sc pamela} package \citep{marsh89-1}. From measurements of
the FWHMs of sky lines we find our observations have a resolution of
approximately 2.5\,pixels (2.2\,\AA) at 8200\,\AA.  
Wavelength calibration of the
extracted spectra was done using only sky emission lines. The
wavelengths of good sky lines were obtained from the atlas of
\citet{osterbrocketal96-1,osterbrocketal97-1}. 37 sky lines were
included, and fitted with a fifth-order polynomial. The rms scatter
around the fit was 0.11\,\AA, so the statistical uncertainty in the
wavelength scale is 0.04\,\AA\ \citep{marshetal94-1}. Finally, the
spectra were flux-calibrated and compensated for telluric absorption
using a spectrum of the standard star Feige\,110.

The Phase\,II design of these observations was to obtain two spectra
per object, separated by at least one night, in order to probe for
radial velocity variations. Due to poor atmospheric conditions,
additional repeat observations were obtained for several targets,
while two objects only got one FORS2 spectrum each.

\subsection{Magellan-Clay/LDSS3}
Follow-up spectroscopy of two PCEBs identified from the FORS2 data,
SDSS\,J1047+0523 and SDSS\,J1414--0123, were obtained over the period
2007 May 17--20 using the Magellan Clay telescope equipped with the
LDSS3 imaging spectrograph. Seeing and transparency were highly
variable. We adjusted the exposure times accordingly, ranging from
500--900\,sec for SDSS\,J1047+0523 and 750--1200\,sec for
SDSS\,J1414--0123.  We used the VPH\_Red grism and OG\,590
order-sorting filter. The detector was an unbinned STA 4k$\times$4k
pixel CCD read out by two amplifiers. A 0.75$^{\prime\prime}$ offset
slit gave a wavelength coverage of 5800--9980\,\AA\, at a reciprocal
dispersion of 1.2\,\AA\,px$^{-1}$. We took flat-field images at each
sky position whilst observing as this was needed to remove the effects
of fringing at redder wavelengths.  The dispersed images were
processed using the same software as for the VLT
observations. Measurements of the FWHMs of the sky lines indicate that
our observational setup gave a resolution of approximately 4\,pixels
(4.8\,\AA) at 8200\,\AA.  Wavelength calibration was done using only sky lines, to
increase observational efficiency. A total of 36 sky lines were fitted
by a fifth-order polynomial. The rms scatter around the fit was
0.25\,\AA, so the statistical uncertainty is 0.09\,\AA. From this work
and other studies we have found that reliable wavelength calibrations
can be obtained from sky lines (see also \citealt{southworthetal06-1}), 
but that the scatter of the fit around
the wavelength solution is quite a bit larger than for arc-lamp
emission lines. In this case the statistical uncertainty is equivalent
to 0.09\,\AA\, so the increased scatter is relatively unimportant.

Flux calibration and telluric line removal was performed using spectra
of the standard star LTT\,3218 obtained during the same observing
run.

\begin{figure*}
\centerline{\includegraphics[angle=-90,width=\textwidth]{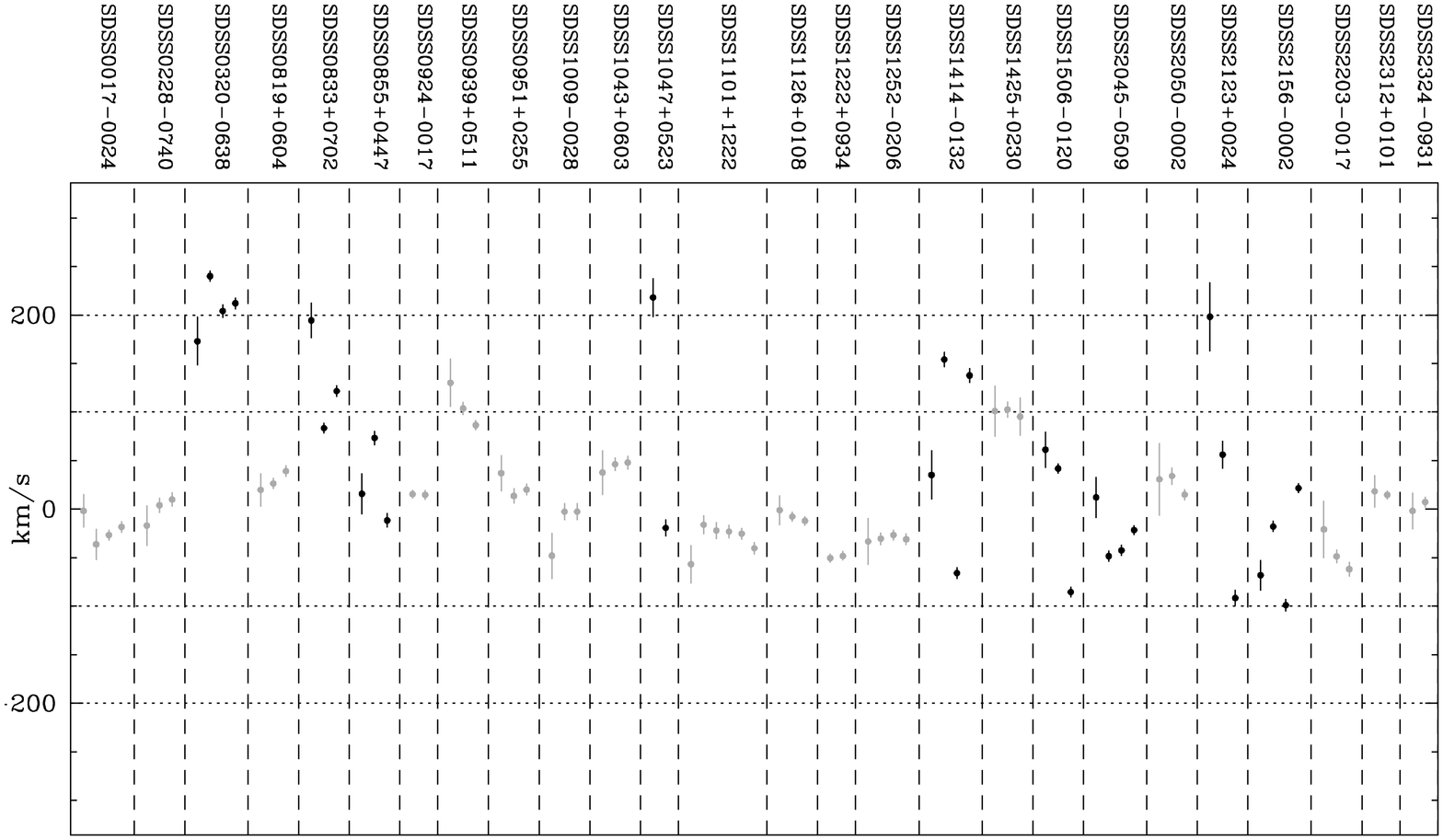}}
\caption[]{\label{f-rvresults} Radial velocities (Table\,\ref{t-rv} of
    our sample 26 WDMS binaries (Table\,\ref{t-targets}), measured
    from the \Lines{Na}{I}{8183.27,8194.81} doublets observed in their
    SDSS and FORS2 spectra. No SDSS radial velocities could be
    determined for for SDSS\,J1252--0206, SDSS\,J1414--0132, and
    SDSS\,J2050--0002. See Fig.\,\ref{f-fitfig} for examples of the
    \Lines{Na}{I}{8183.27,8194.81} fits. The systems with black
    symbols display radial velocity variations at a $3\sigma$ level.}
\end{figure*}

\begin{figure}
\centerline{\includegraphics[width=5cm]{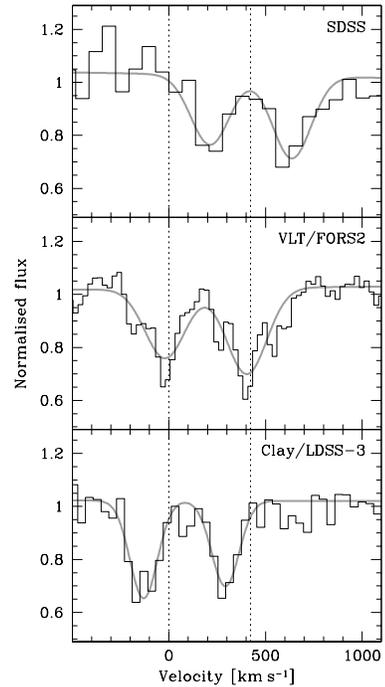}}
\caption[]{\label{f-fitfig} Examples of the double-Gaussian fits to
  the \Lines{Na}{I}{8183.27,8194.81} line profiles in the SDSS,
  VLT/FORS2, and Clay/LDSS3 spectra of SDSS\,J1047+0523.}
\end{figure}

\begin{figure}
\centerline{\includegraphics[angle=-90,width=\columnwidth]{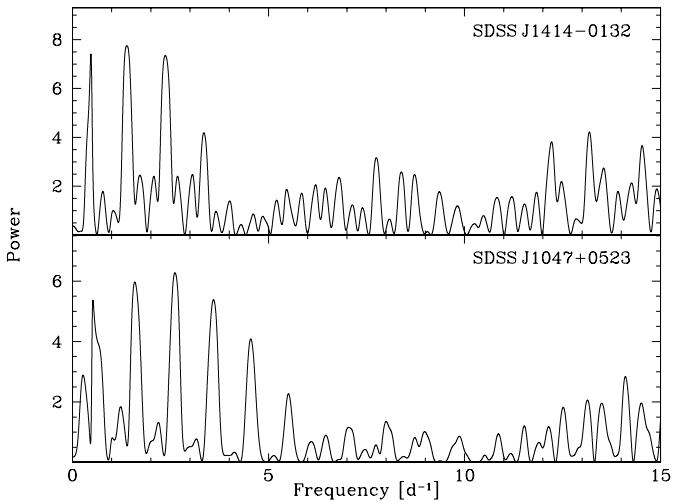}}
\caption[]{\label{f-scargle} Scargle periodograms calculated from the
  \Lines{Na}{I}{8183.27,8194.81}  radial velocities measured in the
  LDSS3 spectra of SDSS\,J1414--0123 and SDSS\,J1047+0523. }
\end{figure}

\begin{figure}
\centerline{\includegraphics[width=\columnwidth]{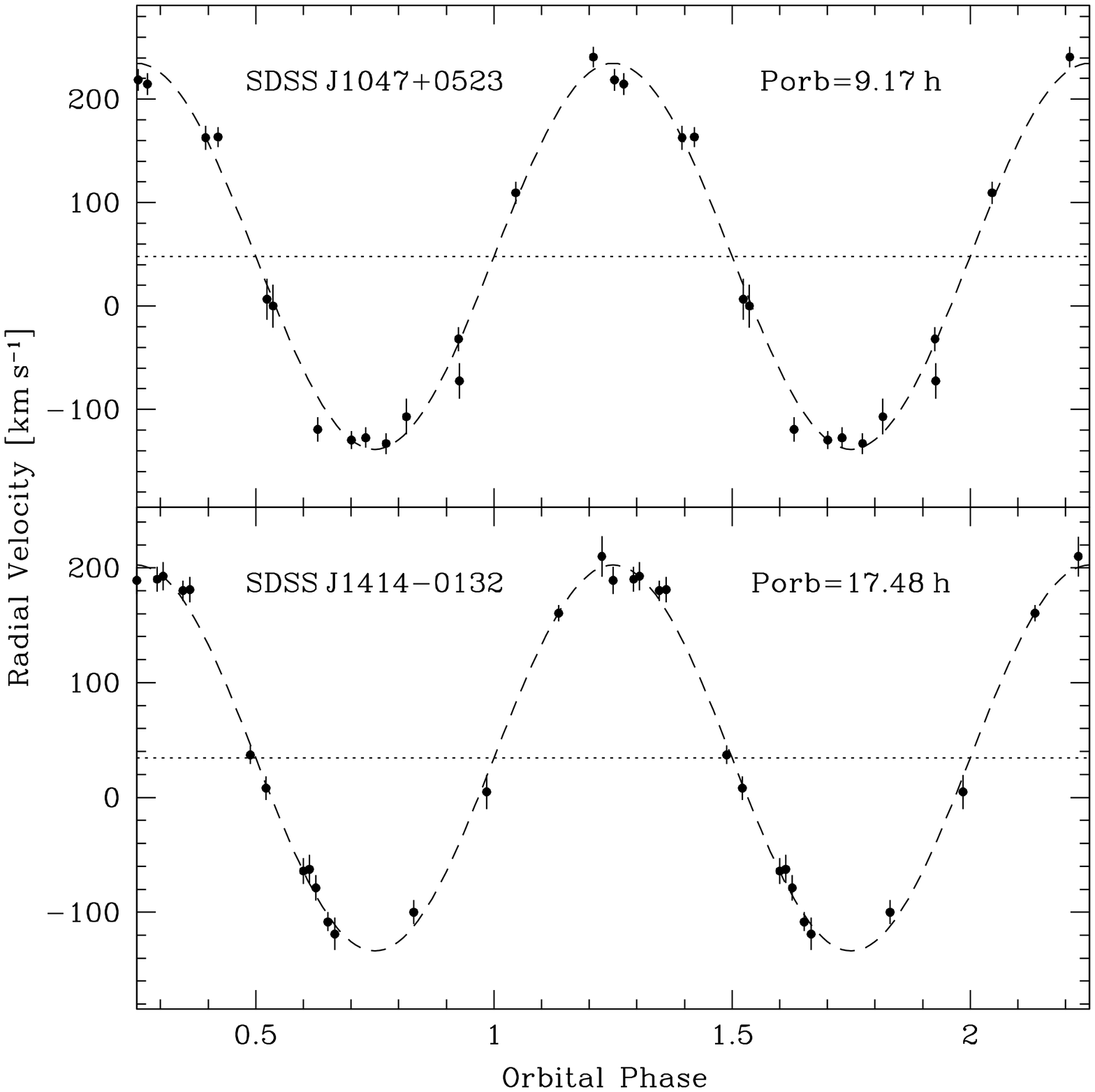}}
\caption[]{\label{f-rvfolded} The \Lines{Na}{I}{8183.27,8194.81}
  radial velocities of the companion stars in SDSS\,J1414--0132 and
  SDSS\,J1047+0523 folded over their respective orbital periods of
  $\Porb=17.48$\,h and $\Porb=9.17$\,h. }
\end{figure}

\section{\label{s-rv}Radial velocity variations and orbital periods}
We have used the strong \Lines{Na}{I}{8183.27,8194.81} absorption
doublet present in the spectra of all our targets to measure the
radial velocities of the main-sequence companion stars. For
this purpose, we fitted the wavelength of the \Ion{Na}{I} doublet
using the sum of a linear slope for the continuum plus two Gaussians
for the absorption lines.  We fixed the separation of the two
Gaussians to 11.54\,\AA\, corresponding to the laboratory value. Free
parameters in the fit were the continuum flux level and slope, the
wavelength of the doublet, the full-width at half maximum (FWHM) of
the lines (both FWHM were fixed to vary together), and the amplitudes
of the individual doublet components.  The fits were carried out using
the \texttt{fit/table} command in \texttt{MIDAS}. Examples of these
fits are shown in Fig.\,\ref{f-fitfig}. We added the uncertainty in
the wavelength calibration in quadrature to the statistical errors in
the wavelength of the \Ion{Na}{I} resulting from the double-Gaussian
fit.

One system, SDSS\,J1222+0934, has an M0V secondary star, which
exhibits no significant \Ion{Na}{I} absorption. For this single
object, we used the same method as above, but fitted three Gaussian
lines of fixed separation and common width to the
\Lines{Ca}{II}{8498.02,8542.09,8662.14} triplet.

We combined the radial velocities obtained from the FORS2 data with
those measured from the DR6 SDSS spectra
(Table\,\ref{t-rv}). Comparing the individual radial velocities of
each object, we define as a strong PCEB candidate all systems which
show a radial velocity variation at a $\ge3\sigma$ level. Applying
this criterion, we identify 9 PCEB candidates among our target sample
of 26 WDMS binaries (Fig.\,\ref{f-rvresults},
Table\,\ref{t-targets}).

The radial velocity variations of the companion stars in
SDSS\,J1414--0132 and SDSS\,J1047+0523 were determined in an analogous
fashion as described above from the Clay/LDSS3 spectroscopy.  We then
calculated Scargle (\citeyear{scargle82-1}) periodograms from the
mean-subtracted radial velocity measurements to establish the
periods of the two WDMS binaries (Fig.\,\ref{f-scargle}). We carried
out sine-fits to the radial velocities using the three highest aliases
in the periodograms as initial value for the frequency, and report
the resulting periods, $\gamma$-velocities and radial velocity
amplitudes, as well as their errors and the reduced $\chi^2$ values, in
Table\,\ref{t-periodaliases}.

\begin{table}
\newcommand{\tb}[1]{\textbf{#1}}
\caption{\label{t-periodaliases} Orbital periods, semi-amplitudes, and
reduced $\chi^2$ from sine-fits to the radial velocity data of
SDSS\,J1047+0523 and SDSS\,J1414--0132 for the strongest three aliases
in the periodograms shown in Fig.\,\ref{f-scargle}. The
best-fit values are set in bold font.}
\begin{flushleft}
\begin{tabular}{lrrrr}\hline\hline
\noalign{\smallskip}
System & \Porb\,[h] & $K_2\mathrm{[km\,s^{-1}]}$ & $\gamma\,\mathrm{[km\,s^{-1}]}$ & $\chi^2$ \\
\noalign{\smallskip}
\hline
\noalign{\smallskip}
SDSS\,J1047+0523 &  15.11$\pm$0.12 & 195$\pm$16 & 20$\pm$11 & 14.4 \\
                 &\tb{9.17$\pm$0.02} &\tb{187$\pm$7} &\tb{45$\pm$5} &\tb{2.8}  \\
                 &   6.64$\pm$0.04 & 181$\pm$22 & 66$\pm$17 & 27.6 \\
SDSS\,J1414--0132 & 51.01$\pm$0.53 & 371$\pm$32 & 44$\pm$12 & 10.8 \\
                  &\tb{17.48$\pm$0.07} &\tb{168$\pm$4}  &\tb{34$\pm$3} &\tb{1.0}  \\
                  & 10.14$\pm$0.09 & 152$\pm$13 & 46$\pm$10 & 13.0 \\
\hline\noalign{\smallskip}
\end{tabular}
\end{flushleft}

\end{table}

The lowest reduced $\chi^2$ values as well as the lowest
relative errors on the fit parameters are found for the periods
corresponding to the highest peaks in the periodograms. We therefore
conclude that the orbital periods of SDSS\,J1414--0132 and
SDSS\,J1047+0523 are $\Porb=17.48\pm0.07$\,h and
$\Porb=9.17\pm0.02$\,h.  Folding the radial velocities over the
neighbouring aliases results in substantially distorted non-sinusoidal
periods.  The LDSS3 radial velocities of SDSS\,J1414--0132 and
SDSS\,J1047+0523 folded over their orbital periods are shown in
Fig.\,\ref{f-rvfolded}.

Below we provide some comments on individual objects. 

\paragraph{SDSS\,J0017--0024:} Two SDSS spectra are available for this
object, with radial velocities that are consistent at a $2\sigma$
level \citep{rebassa-mansergasetal07-1}. We obtained two FORS2
spectra, which agree within their $1\sigma$ errors, confirming this
system as a likely wide binary WDMS. 

\paragraph{SDSS\,J1047+0523:} Only a single VLT/FORS2 spectrum was
obtained of this object. However, combining the radial velocity from
the FORS2 spectrum with that measured from the SDSS spectrum clearly
identifies the system as a close binary. 

\begin{table}[t]
\caption[]{Radial velocities measured from the
\Lines{Na}{I}{8183.27,8194.81} doublet in the ESO/VLT and SDSS
spectra. The radial velocities obtained from SDSS data is indicated by
the HJD set in italics\label{t-rv}. }
\setlength{\tabcolsep}{0.55ex} 
\newcommand{\sd}[1]{\textit{#1}}
\begin{flushleft}
\begin{tabular}{lr@{\hspace*{3ex}}lr}\hline\hline
\noalign{\smallskip}
HJD & RV [$\mathrm{km\,s^{-1}}$] & HJD & RV [$\mathrm{km\,s^{-1}}$] \\
\noalign{\smallskip}
\hline
\noalign{\smallskip}
\multicolumn{2}{c}{\textbf{SDSS\,J0017--0024}}  & \multicolumn{2}{c}{\textbf{SDSS\,J1126+0108}}     \\
2453978.64799       & -26.7$\pm$5.0	        & 2454142.80234      & -7.7$\pm$4.9 	            \\
2453982.67389       & -18.3$\pm$5.5	        & 2454155.62142      & -12.1$\pm$4.7	            \\
\sd{2452518.922000} & -1.8$\pm$16.7   	        & \sd{2451614.80814} & -1.1$\pm$15.4                \\
\sd{2452559.785000} & -36.3$\pm$15.5            & \multicolumn{2}{c}{\textbf{SDSS\,J1222+0934}}     \\
\multicolumn{2}{c}{\textbf{SDSS\,J0228--0740}}  & 2454142.831916 & -50.4$\pm$4.6                    \\
2453983.68162       &    4.0$\pm$8.0   	        & 2454156.875771 & -48.3$\pm$4.5                    \\
2453983.69278       &    9.9$\pm$7.5   	        & \multicolumn{2}{c}{\textbf{SDSS\,J1252--0206}}    \\
-                   &   -16.9$\pm$21.0 	        & 2454142.84647      & -30.4$\pm$6.4   	            \\ 
\multicolumn{2}{c}{\textbf{SDSS\,J0320--0638}}  & 2454169.68656      & -26.7$\pm$5.6  	            \\
2454030.85227      & 240.0$\pm$5.9 	        & 2454185.58421      & -31.1$\pm$6.0                \\
2454086.60870      & 204.1$\pm$7.0 	        & \sd{2451694.70928} &  -33.3$\pm$24.0              \\
2454124.60190      & 212.1$\pm$6.0 	        & \multicolumn{2}{c}{\textbf{SDSS\,J1414--0132}}    \\ 
\sd{2451924.67646} & 172.9$\pm$24.8 	        & 2454142.86192      & 154.2$\pm$7.3 	            \\
\multicolumn{2}{c}{\textbf{SDSS\,J0819+0604}}   & 2454169.71281      & -65.9$\pm$5.7 	            \\
2454030.87927      & 26.4$\pm$5.7  	        & 2454180.89912      & 137.7$\pm$7.8 	            \\
2454055.83644      & 39.2$\pm$6.0  	        & -                  & 35.2$\pm$24.9                \\
\sd{2452963.00286} & 19.8$\pm$17.1 	        & \multicolumn{2}{c}{\textbf{SDSS\,J1425+0230}}     \\ 
\multicolumn{2}{c}{\textbf{SDSS\,J0833+0702}}   & 2453982.48632      & 102.6$\pm$8.4   	            \\ 
2454055.85156      & 83.5$\pm$5.5  	        & 2453983.49473      & 95.3$\pm$19.7   	            \\ 
2454059.82430      & 121.6$\pm$5.9 	        & \sd{2451999.89365} & 101.1$\pm$26.3 	            \\ 
\sd{2452963.98436} & 194.5$\pm$18.3	        & \multicolumn{2}{c}{\textbf{SDSS\,J1506--0120}}    \\ 
\multicolumn{2}{c}{\textbf{SDSS\,J0855+0447}}   & 2453965.51719      & 41.8$\pm$5.3 	            \\ 
2454059.83910      & 73.3$\pm$7.3   	        & 2453982.50834      & -85.4$\pm$5.6	            \\ 
2454060.84131      & -11.7$\pm$7.0  	        & \sd{2452426.79507} & 61.2$\pm$18.6	            \\ 
\sd{2452670.87671} & 15.8$\pm$21.1  	        & \multicolumn{2}{c}{\textbf{SDSS\,J2045--0509}}    \\ 
\multicolumn{2}{c}{\textbf{SDSS\,J0924--0017}}  & 2453978.52649      & -48.4$\pm$5.8 	            \\ 
2454067.842758     & 15.4$\pm$4.5  	        & 2453978.53830      & -42.5$\pm$5.9 	            \\ 
2454086.796712     & 14.7$\pm$5.2  	        & 2453981.61523      & -21.6$\pm$5.2 	            \\ 
\multicolumn{2}{c}{\textbf{SDSS\,J0939+0511}}   & \sd{2452145.71287} & 12.1$\pm$21.2 	            \\ 
2454086.81276      & 103.7$\pm$6.8  	        & \multicolumn{2}{c}{\textbf{SDSS\,J2050--0002}}    \\ 
2454087.84233      & 86.5$\pm$5.4   	        & 2453998.625440     & 34.1$\pm$9.1  	            \\ 
\sd{2452710.81094} & 130.1$\pm$24.9  	        & 2454001.522929     & 15.0$\pm$5.8  	            \\ 
\multicolumn{2}{c}{\textbf{SDSS\,J0951+0255}}   & \sd{2452466.81829} & 30.8$\pm$37.5                \\
 2454086.82795     & 13.6$\pm$7.4   	        & \multicolumn{2}{c}{\textbf{SDSS\,J2123+0024}}     \\  
 2454099.81594     & 20.1$\pm$5.6   	        & 2453978.55706      & 56.1$\pm$14.5 	            \\ 
\sd{2451908.91909} & 37.0$\pm$18.3  	        & 2453981.63252      & -91.6$\pm$7.8 	            \\ 
\multicolumn{2}{c}{\textbf{SDSS\,J1009--0028}}  & \sd{2452523.69331} & 198.2$\pm$35.6	            \\ 
 2454086.84235     &  -2.6$\pm$9.0       	& \multicolumn{2}{c}{\textbf{SDSS\,J2156--0002}}    \\ 
 2454112.83168     &  -2.6$\pm$8.8   	        & 2453992.53760      & -18.0$\pm$5.4 	            \\ 
\sd{2451909.91534} &  -48.0$\pm$23.7  	        & 2453998.64315      & -98.9$\pm$6.4  	            \\ 
\multicolumn{2}{c}{\textbf{SDSS\,J1043+0603}}   & 2454004.55526      & 21.6$\pm$5.1  	            \\ 
2454117.75282      &  46.2$\pm$7.0   	        & \sd{2452075.89855} & -68.1$\pm$15.9 	            \\ 
2454149.70110      &  48.0$\pm$7.2   	        & \multicolumn{2}{c}{\textbf{SDSS\,J2203--0017}}    \\ 
\sd{2452644.01881} &  37.7$\pm$22.9             & 2453978.57300      & -48.7$\pm$6.4  	            \\ 
\multicolumn{2}{c}{\textbf{SDSS\,J1047+0523}}   & 2453982.53714      & -61.9$\pm$7.2  	            \\ 
2454142.77188      & -19.4$\pm$8.9              & \sd{2452173.63031} & -20.9$\pm$29.5 	            \\ 
\sd{2452339.82994} & 218.0$\pm$19.8             & \multicolumn{2}{c}{\textbf{SDSS\,J2312+0101}}     \\ 
\multicolumn{2}{c}{\textbf{SDSS\,J1101+1222}}   & 2453978.59006      & 14.7$\pm$4.7 	            \\ 
2454142.78686      & -16.1$\pm$9.9  	        & \sd{2451811.73493} & 18.3$\pm$16.2	            \\ 
2454155.63799      & -22.0$\pm$8.8  	        & \multicolumn{2}{c}{\textbf{SDSS\,J2324--0931}}    \\ 
2454167.69909      & -23.1$\pm$6.9  	        & 2453963.82187      &  7.3$\pm$5.5    	            \\ 
2454169.67087      & -25.3$\pm$6.0              & \sd{2452203.63065} &  -1.8$\pm$18.7  	            \\ 
2454184.61366      & -40.3$\pm$6.3   	        &\\
\sd{2453119.74299} & -56.8$\pm$19.8             &\\
\hline\noalign{\smallskip}
\end{tabular}
\end{flushleft}
\end{table}

\begin{table}[t]
\caption[]{Radial velocities measured from the
\Lines{Na}{I}{8183.27,8194.81} doublet in the Clay/LDSS3 
spectra of the two PCEB candidates SDSS\,J1047+0523 and
SDSS\,J1414--0132.\label{t-rvldss3}}  
\setlength{\tabcolsep}{0.55ex}
\newcommand{\sd}[1]{\textit{#1}}
\begin{flushleft}
\begin{tabular}{lr@{\hspace*{3ex}}lr}\hline\hline
\noalign{\smallskip}
HJD & RV [$\mathrm{km\,s^{-1}}$] & HJD & RV [$\mathrm{km\,s^{-1}}$] \\
\noalign{\smallskip}
\hline
\noalign{\smallskip}
\multicolumn{2}{c}{\textbf{SDSS\,J1047+0523}}  & \multicolumn{2}{c}{\textbf{SDSS\,J1414--0132}}\\ 
2454237.54059 & 214.7$\pm$10.4    & 2454237.616516 & 190.1$\pm$11.0 \\
2454237.64147 & 0.0$\pm$20.8      & 2454237.625451 & 192.7$\pm$12.1 \\
2454238.48000 &  -127.5$\pm$9.5   & 2454238.587532 & -78.8$\pm$11.2 \\
2454238.51267 &  -107.0$\pm$17.2  & 2454238.736750 & -100.0$\pm$10.5\\
2454238.55507 &  -72.5$\pm$17.1   & 2454239.576845 & 4.8$\pm$14.9   \\
2454238.66262 &  240.7$\pm$9.8    & 2454239.686966 & 160.5$\pm$7.0  \\
2454239.54693 &  6.6$\pm$19.8     & 2454239.752785 & 209.9$\pm$17.2 \\
2454239.58773 &  -119.4$\pm$11.9  & 2454240.498312 & 189.0$\pm$11.8 \\
2454239.61468 &  -129.7$\pm$8.8   & 2454240.569119 & 180.2$\pm$8.8  \\
2454239.64270 &  -133.0$\pm$10.0  & 2454240.579801 & 181.0$\pm$11.0 \\
2454240.46501 &  -31.9$\pm$11.6   & 2454240.671845 & 37.0$\pm$7.7   \\
2454240.51077 &  109.5$\pm$10.5   & 2454240.696022 & 8.1$\pm$10.4   \\
2454240.59002 &  218.7$\pm$10.5   & 2454240.753426 & -64.1$\pm$11.4 \\
2454240.64421 &  162.7$\pm$11.7   & 2454240.762373 & -62.6$\pm$12.0 \\
2454240.65411 &  163.4$\pm$9.6    & 2454240.790358 & -108.4$\pm$7.9 \\
              &                   & 2454240.801040 & -119.1$\pm$13.8\\
\hline\noalign{\smallskip}
\end{tabular}
\end{flushleft}
\end{table}

\section{\label{s-dis}Discussion}

\subsection{\label{s-porblimits} Orbital period limits}

As in \citet{rebassa-mansergasetal07-1} we determine an upper limit
for the orbital period based on the measured radial velocity
variations (Table\,\ref{t-rv}).  The orbital period as a function of the stellar masses,
the binary inclination ($i$), and the radial velocity amplitude
($K_{\mathrm{sec}}$) of the secondary follows from Kepler's 3rd law:
\begin{equation}
\frac{(\Mwd \sin i) ^3}{(\Mwd + \Msec)^2}=\frac{\Porb K_{\mathrm{sec}}^3}{2\pi\,G}
\end{equation} 
Using the stellar parameters estimated in Sect.\,\ref{s-modelling}
from fitting the SDSS spectrum, assuming $i=90^\circ$, and assuming
the largest radial velocity difference measured for one particular
system to be equal $2\,K_{\mathrm{sec}}$, clearly gives an upper limit
on $\Porb$.  This method was used in
\citet{rebassa-mansergasetal07-1}.  Here we slightly improve this
approach by using also the information provided by the times of
observations.  If the time delay between two spectra ($\Delta t$) is
shorter than half the considered orbital period, the maximal radial
velocity difference that could have been observed writes
$2K_{\mathrm{sec}}\sin(\pi\Delta t/2\Porb)$
and we modify Eq.\,(1) accordingly.  
Upper limits on the orbital period for the $9$ PCEBs obtained with this
method are given in Table\,\ref{t-targets}.

\subsection{The fraction of PCEBs among SDSS WDMS binaries}

The first result of our PCEB survey will be the relative
number of PCEBs among SDSS WDMS binaries. 
According to the derived limits on the orbital periods and the masses
given in Table\,1, the $9$ close WDMS binaries identified in this paper 
clearly experienced a phase of mass transfer during the post main sequence 
evolution of the progenitor of the white dwarf. 
As mentioned in the Introduction, according to current WDMS binary population
studies (with many uncertainties involved), the majority of all WDMS that 
formed through mass transfer
interactions are expected to be PCEBs and only a small fraction of close
systems may form through dynamically stable mass transfer. 
For the $9$ PCEB candidates identified among our $26$ WDMS systems 
dynamically stable mass transfer appears to be extremely unlikely as the white
dwarf masses are rather high and the secondary spectral types are rather
late \citep[see][for more details]{willems+kolb04-1}. We therefore state to
have identified 9 PCEBs.

In the first paper of this series
\citep{rebassa-mansergasetal07-1} multiple SDSS spectroscopy has been
used to identify 18 PCEB candidates among 101 WDMS binaries from the SDSS.  Due
to the low spectral resolution of SDSS spectra and the fact that in
most of those 101 WDMS binaries only two SDSS spectra are available, the
resulting PCEB fraction of $\sim15\%$ certainly has to be considered
as a lower limit on the true value.  In this paper we identified 9
PCEBs among 26 SDSS WDMS binaries which gives a PCEB fraction of
$35\pm12\%$, which is --~as expected~-- substantially higher than our
previous lower limit. The increased PCEB detection rate is probably
the result of both, the higher resolution of VLT/FORS2 and the fact
that for most systems 3--4 spectra are available.

\begin{figure}
\centerline{\includegraphics[angle=270,width=\columnwidth]{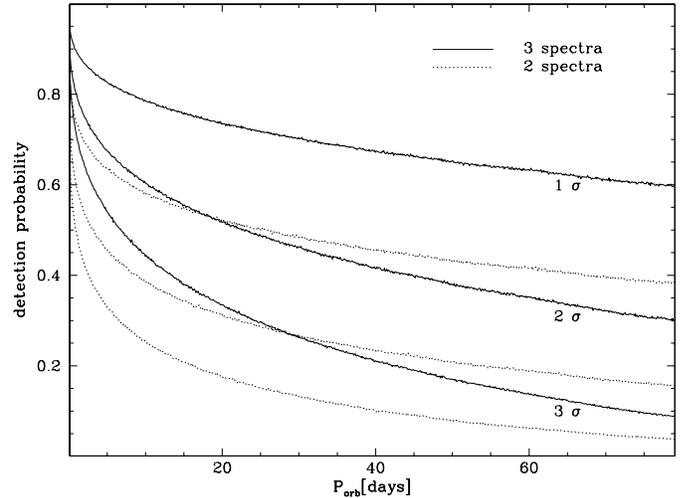}}
\caption[]{\label{f-depro}Detection probability versus orbital period for $2$
and $3$ spectra taken randomly in one semester but not in the same night. 
We here assumed $\Mwd=0.6\Msun$ and $\Msec=0.3\Msun$. The
$1-3\sigma$ labels refer to the detection level of radial velocity
variations. Apparently, using our $3\sigma$ criterion to select strong 
PCEB candidates for further follow-up observations leads to final a sample
that is biased toward shorter orbital periods. 
However, only $3$ spectra are needed to identify $15-30\%$ of the  
PCEBs with longer orbital periods ($\Porb\gappr\,20\,$days) and our strategy 
can therefore certainly be used to also constrain the orbital period
distribution of long orbital period PCEBs.}
\end{figure}

However, also in the case that more than 2 spectra are available exists 
a certain probability to sample similar orbital phases and also the
observations presented here have a finite spectral resolution. In the
context of our project on PCEB studies, it is important to understand
possible selection effects in the established PCEB sample. We
therefore developed a Monte-Carlo code to estimate the detection
probability among SDSS WDMS binaries as a function of orbital
period. For the purpose of this specific paper, we limit our analysis
to the following assumptions: a mean radial velocity resolution of
$6$\,km/s, stellar masses of $\Mwd=0.6\,\Msun$ and $\Msec=0.3\,\Msun$,
and uniformly distributed inclinations. For an assumed
orbital period we randomly select the inclination, orbital phase, and
times of observations. The only constraint on the latter being that
the time differences should exceed $12$\,h and that all observations
are performed in one semester. The randomly selected values are then
used to calculate radial velocity differences. Repeating this
exercise $10^4$ times and counting the number of cases with a
$3\sigma$ (alternatively $1$ or $2\,\sigma$) radial velocity
variation gives the probability of detecting a PCEB with the assumed
orbital period.  Fig.\,\ref{f-depro} shows the results we obtain
assuming that 2 or 3 spectra have been taken in one semester. In both
cases the detection probability of different significance levels of
radial velocity variations is shown. Of course, our strategy leads to
a certain bias towards shorter orbital periods (and higher
inclinations) because strong radial velocity variations are more
likely to be detected in these systems.  However, with three randomly
taken VLT/FORS2 spectra our $3\sigma$ criterion is still $\sim15-30\%$
sensitive to PCEBs with periods in the range of
$\sim20-60$\,days. PCEBs with orbital periods longer than $70$ days
may hide among the systems with weak ($2\sigma$) radial velocity
variations, and will require additional spectroscopy to be
unambiguously identified.  In general, Fig.\,\ref{f-depro} clearly
shows that the fraction of PCEBs among SDSS WDMS binaries of $35\pm12\%$
obtained here still represents a lower limit. However, as mentioned
in the Introduction, the relative number of PCEBs may be a function of
the spectral type of the secondary.

\subsection{Dependence on the secondary star spectral type}

Recently, \citet{politano+weiler06-1} suggested a test for the
disrupted magnetic braking scenario, i.e. the hypothesis that magnetic
braking is not active in PCEBS which contain a fully convective
companion star.  Their calculations predict that the relative number
of PCEBs among WDMS binaries should drastically increase for
secondary star masses below $\Msec=0.35\Msun$. Our results are
consistent with that prediction. Among $9$ WDMS binaries with spectral
type M3 or earlier we find only one strong PCEB candidate while the
PCEB identification rate among WDMS binaries containing later
secondaries (M4--M6) is 8/17.  
According to the Poisson statistic, the
relative numbers of PCEBs among SDSS WDMS binaries are $0.111\pm0.111$
(for M0--M3) and $0.471\pm0.166$ (for M4--M6).  
Combining these values we obtain a significance of $1.8\sigma$ 
for the hypothesis of having two different distributions. 
Interestingly, a potential difference between the secondary star spectral
types in close and wide WDMS binaries was also suggested by
\citet[][]{farihietal05-1}, who found, subject to small 
number statistics, that the median secondary spectral type of close
binaries is one spectral class later than that of the wide
binaries. While keeping in mind the well known problems resulting from low 
number statistics, these results may represent first indications for a 
discontinuity in the relative number of PCEBs among WDMS binaries 
as predicted by \citet{politano+weiler06-1}. However, further
observations are certainly needed to confirm or disprove the observed
tendency.

\subsection{Predicted evolution of SDSS\,J1047+0523 and SDSS\,J1414--0132}

As shown by \citet{schreiber+gaensicke03-1} knowing the 
mass and effective temperature of the white dwarf, its orbital period, and 
at least reasonable estimates for the mass and the radius 
of the secondary one can determine the age of the PCEB, the orbital period of
the future semi-detached system (i.e. a cataclysmic variable) and the 
time the PCEB evolution takes for a given system. 

To that end, we here first follow \citet{schreiber+gaensicke03-1} by
using the white dwarf cooling tracks of \citet{wood95-1} to determine
the white dwarf cooling ages of SDSS\,J1047+0523 and SDSS\,J1414--0132.  Using
the white dwarf parameters given in Table\,1 and interpolating the cooling
tracks we get $t_{\mathrm{cool}}\sim2.1\pm1\times10^{8}$\,yrs and
$t_{\mathrm{cool}}\sim3.9\pm1.1\times10^{8}$\,yrs 
for SDSS\,J1047+0523 and SDSS\,J1414--0132 respectively.
  
The determination of the mass and the radius of the secondary is much
more challenging as the inclination of these two systems is
unconstrained. We use the empirical spectral type-radius-mass (Sp-R-M)
relation derived in Paper\,I (Sect.\,3.4) to
estimate values of the mass and radius of the secondary.  Apparently,
the scatter around this empirical relation is quite large
(Paper\,I, Fig.\,7) and therefore the obtained values should be 
generally considered as uncertain. However, in the case of 
SDSS\,J1414--0132 and SDSS\,J1047+0523 the distance estimates based 
on the white dwarf ($d_{\mathrm{WD}}$) and those derived using the 
empirical Sp-R-M relation agree quite well which suggests that 
the obtained values are
plausible (for a detailed discussion of the spectral
type-mass-radius relation see Paper\,I). As both
secondaries are of the spectral type M5 the empirical relation gives
$\Msec=0.216\Msun$ and $\Rsec=0.258\Rsun$. For our theoretical
analysis we adopt rather broad ranges of possible values to account
for the described uncertainties, i.e.  $\Msec=0.20-0.23\Msun$ and
$\Rsec=0.2-0.3\Rsun$.

\begin{figure}
\centerline{\includegraphics[angle=270,width=\columnwidth]{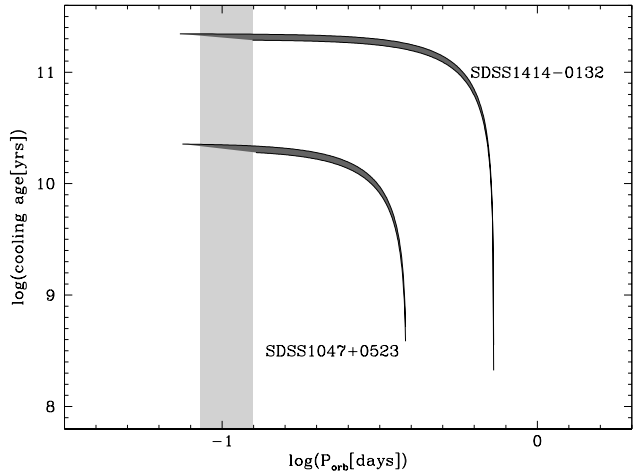}}
\caption[]{\label{f-evol} The PCEB evolution of SDSS\,J1047+0523 and
SDSS\,J1414--0132.  Starting at their current orbital periods of
$\Porb=17.48$\,h and $\Porb=9.17$\,h the systems evolve towards
shorter orbital period due to angular momentum loss by gravitational
radiation. The companion stars in SDSS\,J1047+0523 and
SDSS\,J1414--0132 will fill their Roche-lobes in
$1.7\pm0.3\times\,10^{10}$\,yrs and $1.8\pm0.2\times10^{11}$\,yrs,
respectively, and turn these PCEBS into cataclysmic variables.  This
will happen at very similar orbital periods, $0.090\pm0.013$\,days for
SDSS\,J1047+0523 and $0.085\pm0.015$ for SDSS\,J1414--0132.}
\end{figure}

According to the standard scenario of CV evolution, magnetic braking
gets disrupted when the secondary stars becomes fully convective
\citep[e.g.][]{verbunt+zwaan81-1} and, hence, gravitational radiation
is the only angular momentum loss mechanism acting in systems with a
low mass ($\Msec\lappr\,0.3\Msun$) secondary star.  The evolution of
SDSS\,J1047+0523 and SDSS\,J1414--0132 according to this scenario and
taking into account the uncertainties discussed above, is shown in
Fig.\,\ref{f-evol}. Both systems will become CVs in or slightly below
the $2-3$h orbital period gap and for both systems the evolution will
take longer than a Hubble-time. Therefore, according to
\citet{schreiber+gaensicke03-1}, they are not representative for
progenitors of the current CV population.

\section{Summary} 

We have obtained ESO-VLT/FORS2 spectroscopy of 26 WDMS binaries found
in SDSS.  Clear radial velocity variations led to the identification
of 9 strong PCEB candidates, which gives a hit-rate of $35\pm\,12\%$.  We
analysed the PCEB detection probability of our strategy and conclude
that (although a bias towards short orbital period systems is
inevitable) our survey is sensitive to the predicted long orbital
period distribution of PCEBs.  While we find 8 PCEB candidates among
17 systems with M4 or later secondary stars only one PCEB candidate
could be identified among 9 WDMS binaries containing earlier secondary
stars. Although suffering from low number statistics, one may
interpret this results as a first indication for a discontinuity in
the dependence of the relative number of PCEBs among WDMS binaries on
the spectral type of the secondary.

We determined the orbital periods for two of the PCEBs identified
here, SDSS\,J1047+0523 and SDSS\,J1414--0132, from
Magellan-Clay/LDSS3 spectroscopy to be to be $9.17$\,h and $17.48$\,h
respectively.  Both systems will enter the semi-detached CV state in
or below the orbital period gap. This evolution will last longer than
the Hubble time and both systems are therefore different than the
progenitors of the current CV population.

\begin{acknowledgements}
MRS acknowledges support from FONDECYT (grant 1061199), DIPUV (project 35),
and the Center of Astrophysics in Valparaiso.
\end{acknowledgements}


\end{document}